\documentstyle[12pt]{article}
\begin{document}
\begin{center}
{\bf A Superconnection for Riemannian Gravity as Spontaneously Broken
SL(4,R) Gauge Theory}\\
\vspace{.2in}
Yuval Ne'eman\\
Sackler Faculty of Exact Sciences\footnote{Wolfson Distinguished Chair in
Theoretical Physics}, Tel-Aviv University, Israel 69978 \\
and\\
Center for Particle Physics, University of Texas, Austin, Texas 78712 \\
\vspace{.2in}
{\bf Abstract}
\end{center} 
\vspace{.2in}
A {\it superconnection} is a supermatrix whose even part contains the
gauge-potential one-forms of a local gauge group, while the odd parts contain
the (0-form) Higgs fields; the combined grading is thus odd everywhere.
We demonstrate that the simple supergroup ${\overline P}(4,R)$ (rank=3) in Kac'
classification (even subgroup $\overline {SL}(4,R)$) provides for the most
economical spontaneous breaking of $\overline{SL}(4,R)$ as gauge group, leaving
just local $\overline{SO}(1,3)$ unbroken. As a result, post-Riemannian SKY gravity
yields Einstein's theory as a low-energy (longer range) 
effective theory. The theory is renormalizable and may be unitary.

\vspace{.3in}
\noindent
{\bf 1. Superconnections and the Electroweak $SU(2/1)$ as a model.}

The {\it superconnection} was introduced by Quillen [1] in Mathematics.
It is a supermatrix, belonging to a given supergroup $S$, valued over elements
belonging to a Grassmann algebra of forms. The even part of the supermatrix
is valued over the gauge-potentials of the even subgroup $G\subset S$,
({\it one-forms} $B^{a}_{\mu} dx^{\mu}$ on the base manifold of the bundle,
realizing the ``gauging" of $G$). The odd part of the supermatrix,
representing the quotient $S/G=H\subset S$, is valued over {\it zero-forms}
in that Grassmann algebra, physically the Higgs multiplet $\Phi_{H}(x)$,
in a spontaneously broken $G$ gauge theory;
$\phi(x)\in\Phi_H(x),\hfil\break
 <0|\phi(x)|0>\neq 0$, thus leaving only a subgroup
$F\subset G, [F,\phi]=0$ as the low-energy residual local symmetry.
In quantum treatments which are set to reproduce geometrically the ghost
fields and BRST equations [2],
the Grassmann algebra is taken over the complete
bundle variable. For simplicity in the presentation,we shall leave out that
aspect in this work.

The first physical example of a superconnection preceeded Quillen's theory.
This was our $SU(2/1)$ (supergroup) proposal for an algebraically irreducible
electroweak unification [3,4]. Lacking Quillen's generalized formulation,
the model appeared to suffer from spin-statistics interpretative complications
for the physical fields. The structural $Z_2$ {\it grading} of Lie
superalgebras, as previously used in Physics (i.e. in {\it supersymmetry})
corresponds to the grading inherent in quantum statistics, i.e. to Bose/Fermi
transitions, so that invariance under the supergroup represents
{\it symmetry between bosons and fermions}. Here, however, {\it though
the superconnection itself does fit the quantum statistics ansatz, this is
realized through the order of the forms in the Grassmann algebra, rather
than through the quantum statistics of the particle Hilbert space}!
(both the $W^{\pm}_{\mu}, Z^{0}_{\mu}, A_{\mu}$ on the one hand, and the
Higgs field $\Phi_{H}$ on the other, are bosons). Moreover, the  matter
fields' (leptons or quarks) Hilbert space is $Z_2$-graded by
{\it fermion chiralities}. The internal quantum numbers set by the
$SU(2/1)$ assignments do display a perfect fit with the phenomenology
(including the provision that integer charges - leptons - come in three
states (e.g. $[\nu_{L},e_{L}/e_{R}]$), whereas fractional charges - quarks -
come in four (e.g. $[u_{R}/u_{L},d{L}/d_{R}]$), but they obviously do not lend
themselves to a quantum statistics grading ansatz.

Two important sequels followed at that early stage.
On the one hand, it was noted [5] that the method appears to apply to a
large set of spontaneously broken symmetries, global or local.
The example of the Goldstone-Nambu breaking of (global) {\it chiral}
$SU(3)\otimes SU(3)$ was treated in detail in ref. [5], the relevant
supergroup being the $Q(3)$ of Kac' classification [6] of the simple
superalgebras. This is an exceptional supergroup we had encountered 
earlier [7], precisely because of its physical relevance.
The second development was an improved understanding of both the physics
and the mathematics of the juxtaposition of the two graded systems -
on the one hand, the supergroup as represented by its supermatrices and on
the other hand the Grassmann algebra over which it is valued [8].
Tied to the traditional Yang-Mills derivation, however, our Grassmann-even
elements, in the group-odd sector of the superconnection, started with
two-forms, thus missing the desirable zero-forms, which Quillen could freely
postulate\footnote{Our 1982 solution nevertheless did include a fitting
scalar field, within the extended ``ghost" system corresponding to the
forms being taken over the entire bundle}.
Our final treatment, with S. Sternberg [9], availed itself of the
advantage deriving from Quillen's generalized formalism, also applying the
method to a further unification [10], including QCD and a ($2k$)
generations structure, using $SU(5+k/1)$.   

One last `technical' point relates to the multiplication of supermatrices.
To stay within the axioms of matrix multiplication, terms in the product have
to take appropriate signatures:
$$
\left[ \left( \begin{array}{cc}
A & B\\ C & D
\end{array} \right)\right]\ .
\left[ \left( \begin{array}{cc}
A' & B' \\ C' & D' \end{array} \right)\right] =
$$
$$
\left[ \left( \begin{array}{cc}
A \wedge A'+ (-1)^b B\wedge C' & (-1)^a A\wedge B'+ B\wedge D' \\
C\wedge A'+ (-1)^d D\wedge C' & (-1)^c C\wedge B'+ D\wedge D'
\end{array}\right)\right]
\eqno(1)
$$
where $n=a,b,c,d$ are the respective orders of the n-forms $A,B,C,D$ in the
Grassmann algebra. 
  
The next installment came from Connes' noncommutative geometry (NCG),
generalizing to discrete geometries some geometrical concepts (such as
distances) till then defined only for continuous spaces.
Connes and Lott [11] used the new formalism to reproduce the
{\it electroweak} theory, providing it with a geometric derivation: the
base manifold is $Z_2\otimes M^{3/1}= M_L\oplus M_R$, where $Z_2$ is a
discrete space containing just two points $L,R$ representing {\it chiralities}
and $M^{3/1}$ is Minkowski spacetime.\footnote{The authors of
ref. [11] worked on Euclidean $M^4$, instead of $M^{3/1}$ for
technical reasons.}
NCG defines a space by the functions
and Hilbert space states living on it and the operators acting on that
Hilbert space. Here, parallel-transport within $M_L$ (or within $M_R$)
is performed by $D=d+B(G)$, with $B$ standing for the relevant gauge
potentials and $G=SU(2)_{W}\times U(1)_{Y}$. Moving, however, from a state
sitting over a point in $M_{L}$ (say $\nu^{e}_{L}(x)$) to one sitting over a
point in $M_{R}$ (say $e_{R}(x)$) requires a scalar ``connection"
$\Phi_{H}(x)$. In this case, its $G$ quantum numbers are entirely fixed by
the matter fields' selected assignments; this includes the Lorentz scalar
nature of $\Phi_{H}$, due to its having to relate e.g.
$e^{-}_{L}$ to $e^{-}_{R}$ in a Yukawa term (the $\gamma^{0}$ is provided
for by the Dirac operator, acting on the Hilbert space, in the definition
of the noncommutative geometry). 

The link with our $SU(2/1)$ superconnection was provided by Coquereaux,
Scheck and coworkers of the ``Marseilles-Mainz group" [12-16]. 
They found that by slightly modifying the Connes axioms, su(2/1)
{\it emerges naturally as the superalgebra of the form-calculus over the
discrete} $Z_2$ {\it of the chiralities, while the superconnection for the
product space} $Z_2\otimes M^{3/1}$ is an SU(2/1) group-element.
{\it This also finally exorcizes the apparent difficulty with the
non-spin-statistics grading of the matter fields and explains how the
grading can be related to chiralities instead.}
Moreover, the parallel-transport operator is found to require an additional
``matrix derivative" $\delta_H$, relating ``twin" states in $M_L$ and $M_R$,
such as $e_L$ and $e_R$, etc.. (this is the role of $\beta$ in the Dirac
$\gamma^\mu$ calculus). With this additional term, the curvature-squared
Lagrangian $\hat{R}\wedge {}^{*}\hat{R}$ for $S=SU(2/1)$
contains the complete Weinberg-Salam Lagrangian. Indeed,
$\hat{R}=R_G+(1/2)\{\Phi_H,\Phi_H\}+D_G\Phi_H+\delta_H\Phi_H$,
with $R(G)=dB+(1/2)B\wedge B$. In squaring, the second term in $\hat{R}$
provides for the $\lambda\Phi^{4}$ and the fourth provides for the negative
mass-squared piece of the Higgs potential. 
\vspace{.3in}

\noindent
{\bf 2. Riemannian Gravity as deriving from a broken SKY Affine Gravity}

The interst in deriving Einstein's Riemannian theory through the spontaneous
symmetry breakdown of a non-Riemannian theory stems from quantum
considerations. First, the quantization of gravity implies spacetime
quantization at Planck energies (where the Compton wavelength is also the
Schwarzschild radius, $(h/2\pi mc)=2Gm/c^2)$.
This quantization, in  itself, represents a departure from Riemannian
geometry. Secondly, the addition in the Lagrangian of terms quadratic in the
curvatures renders the theory finite (the new terms dominate at high-energy
and are dimensionless in the action); however, it is nonunitary, due to the
appearance of $p^{-4}$ propagators. These are present because of the
Riemannian condition $Dg_{\mu\nu}=0$, relating the connection $\Gamma$ to
the metric $g_{\mu\nu}$ (the Christoffel formula).
Thus $\Gamma\simeq\partial g$ and
$R=d\Gamma+(1/2)\Gamma\Gamma\simeq(\partial)^2 g + (\partial g)^2$ and
$R^2$ will involve $p^4$ terms in momentum space and thus $p^{-4}$
propagators. These can then be rewritten as differences between two
S-matrix poles - one of which is then a ghost, due to the wrong sign of its
residue. It seems therefore worth trying to {\it reconstruct gravity
so that the Riemannian condition will only constrain the low-energy end of
the theory, as an effective result in that regime.}
The high-energy theory, i.e. prior to symmetry breakdown, should have as
its anholonomic (gauge) group the metalinear
$\overline{SL}(4,R)$.\footnote{The conformal
$SU(2,2)=\overline{SO}(4,2)$ or the
super-deSitter $OSp(4/1)$ of McDowell-Mansouri would still
be Riemannian.}
We have investigated a model [17-20] based on either
$\overline{SL}(4,R)$ or
$\overline{GL}(4,R)$, containing the Stephenson-Kilmister-Yang (SKY)
Lagrangian [21-23] plus a term linear in the curvature, and proved the
Yang-Mills-like renormalizability and BRST invariance of the quantum
Lagrangian. Whether the theory is unitary is not known at this stage,
due to the presence of a $p^{-4}$ term as input in the gauge-fixing term
of the quantum Lagrangian${}^{[19-20]}$.

In such theories, (a) the $G=\overline{SL}(4,R)$-invariant
$\hat{R}(G)\wedge{}^{*}\hat{R}(G)$ SKY Lagrangian has to have its
symmetry broken by a Higgs field corresponding to an $\overline{SL}(4,R)$
multiplet containing a Lorentz-scalar component, to ensure that
$F=\overline{SO}(1,3)$.  In the algebraic structure we use (the superalgebra of
$S=\overline{P}(4,R)$), this includes a metric-like symmetric tensor ($a,b=0,..3$
are anholonomic indices supporting the local action of $S$ and its subgroups)
$\Phi_{\{ab\}}(x)$, and the Lorentz scalar is given by
$\phi=\Phi_{ab}\eta^{ab}$, where $\eta_{ab}$ is either the trace
(for Euclidean signature situations) or the Minkowski metric.
Thus $<0|\phi(x)|0>\neq 0$. (b) Those components of the connection
$\Gamma^a_b(x)$ which serve to gauge
$G/F=\overline{SL}(4,R)/\overline{SO}(1,3)$
should acquire masses in the spontaneous breakdown procedure.
As in the electroweak case, we should have in the Higgs multiplet, components
which - in the Unitary gauge - will have become the longitudinal (spin)
components of the (now massive) $G/F$ elements of the connection.
In our construction, these are precisely the 9 components of
$\Phi_{\{ab\}}$, {\it after removal of the trace (or Minkowski-trace)}.
(c) Any remaining components of $\Phi_H(x)$ should acquire masses and exist
as free particles. In the $\overline{P}(4,R)$ model,
$\Phi_H(x)=\Phi_{\{ab\}}(x)\oplus\Phi_{[ab]}(x)$, i.e. there is, in addition,
an antisymmetric field $\Phi_{[ab]}(x)$, which indeed acquires a
Planck-scale mass. 

\vspace{.3in}
\noindent
{\bf 3. The simple superalgebra $p(4,R)$.}

The defining representation of the generating superalgebra of the
$\overline{P}(4,R)$ supergroup is an $8\times 8$ matrix, divided into quadrants.
$I$ and $IV$ carry the $sl(4,R)$ algebra, with $I$ in the covariant
representation $\Sigma^{\tilde a}_{\tilde b}$ ($a,b=0,1,2,3$ and the tildes
indicate tracelessnes $tr\Sigma^{\tilde a}_{\tilde b}=0$) and $IV$ in the
contravariant, i.e. $IV= - I^T$ (${}^{T}$ indicates transposition).
In the off-diagonal quadrants, $II =\Sigma^{\{a}_{b\}}$ carries the 10
symmetric matrices of $gl(4,R)$ and $III=\Sigma^{[a}_{b]}$ carries its 6
antisymmetric matrices. There are thus altogether 31 generators, of which
15 $Q^{\tilde a}_{\tilde b}$ are even, representing the action of $sl(4,R)$ and
16 $N^a_b$ are odd, of which 10 are the symmetric $N^{+}=T$,
and 6 the antisymmetric $N^{-}=M$,exhausting the set of generators of
$gl(4,R)$ (we use the notation of ref. [24], i.e. the $T,\ M$ are the shears
and Lorentz generators, respectively). We shall also have occasion to use
the nonsimple completion $\overline{gP}(4,R)$ in which the $\overline{SL}(4,R)$
even subgroup is completed to $\overline{GL}(4,R)$, without any change in the
${\overline P}(4,R)$ itself.

The simple superalgebra is thus given as,
$$ \left(
\begin{array}{cc}
I   &  II \\
III &  IV 
\end{array} \right)
\begin{array}{rcccl}
I   &=& \Sigma^{\tilde a}_{\ \tilde b}& =& \uparrow Q_{\tilde a\tilde b} \\
II  &=&\Sigma^{\{a}_{\ b\}}& = &T_{ab} \\  
III &=& \Sigma^{[a}_{\ b]} &=& M_{ab} \\
IV &=& - (\Sigma^{\tilde a}_{\ \tilde b})^{T} & = &
\downarrow Q_{\tilde a\tilde b}  
\end{array}
\eqno(2)
$$
and
$$
\begin{array}{rcl}
Q_{ab} & := &(\Sigma^{\tilde a}_{\ \tilde b})_{I}\oplus -
(\Sigma^{\tilde a}_{\tilde b})^T_{IV} \\
N_{ab}^{+} & := &(\Sigma^{\{a}_{b\}})_{II} \\
N_{ab}^{-} & := &(\Sigma^{[a}_{b]})_{III}  
\end{array} \eqno(3)
$$

To formulate the super-Lie bracket, we choose to replace the two-index
(vector) notation by a single (matrix) index, as in SU(2) or SU(3) usage.
We select an SU(4) basis ($4\times 4,``\nu$" matrices) in which the $i=1..8$
correspond to setting the SU(3) $\lambda_{i}$ matrices in the upper left-hand
corner of the $\nu$ matrix with that index and define similar matrices for
the rest. Since we are dealing with $SL(4,R)$ rather than $SU(4)$, we have
to multiply the real matrices by $\sqrt{-1}$, thus making these generators
noncompact. With $\sigma_i$ denoting the Pauli matrices, and
$[\sigma_i]_{1,2}$ denoting a $\sigma_1$ matrix placed in the $[1,2]$ rows
and columns of the $\nu$ matrix, we have a basis,
$$
\begin{array}{rcccl}
\nu_{1} &=& i\lambda_{1}&= & i[\sigma_{1}]_{1,2}\\
\nu_{2} &= & \lambda_{2}&= & [\sigma_{2}]_{1,2} \\
\nu_{3} &=& i\lambda_{3}&= & i[\sigma_{3}]_{1,2} \\
\nu_{4} &=& i\lambda_{4}&= & i[\sigma_{1}]_{1,3} \\
\nu_{5} &= & \lambda_{5}&= &  [\sigma_{2}]_{1,3} \\
\nu_{6} &=& i\lambda_{6}&= & i[\sigma_{1}]_{2,3} \\
\nu_{7} &= & \lambda_{7}&= &  [\sigma_{2}]_{2,3} \\
\nu_{8} &= & i\lambda_{8}&=& (i/\sqrt{3})diag(1,1,-2) \\
\nu_{9} &= &     --             &=& i[\sigma_{1}]_{1,4} \\
\nu_{10}&= &     --             &=&  [\sigma_{2}]_{1,4} \\
\nu_{11}&= &     --             &=& i[\sigma_{1}]_{2,4} \\
\nu_{12}&= &     --             &=&  [\sigma_{2}]_{2,4} \\
\nu_{13}&= &     --             &=& i[\sigma_{1}]_{3,4} \\
\nu_{14}&= &     --             &=&  [\sigma_{2}]_{3,4} \\
\nu_{15}&= &     --             &=& (i/\sqrt{6})diag(1,1,1,-3) 
\end{array}
\eqno(4)
$$

Using the definition of the $f_{ijk}$ (totally antisymmetric) and $d_{ijk}$
(totally symmetric) coefficients of $su(3)$, generalized to $su(4)$ and
corrected by the factors $\sqrt{-1}$ for the symmetric matrices in the
$su(4)$ basis when changing to $sl(4,R)$ as indicated above, we get 
coefficients $\hat{f}_{ijk}$ and $\hat{d}_{ijk}$ whose symmetry properties
are thus reduced to the first two indices only. 
We can now  write the Lie superbrackets as, 
$$
\begin{array}{rcl}
[Q_i, Q_j] &=& 2i \hat{f}_{ijk} Q_k \\
\lbrack Q^A_i, N^+_j\rbrack &=& 2i\hat{f}_{ijk} N^+_k \\
\lbrack Q^A_i, N^+_0\rbrack &=& 0 \\ 
\lbrack Q^T_i, N^+_j\rbrack &=& 2\hat{d}_{ijk} N^+_k \\
\lbrack Q^T_i, N^+_0\rbrack &=& 2i N^+_i \\                  
\lbrack Q^T_i, N^-_j\rbrack &=& 2i\hat{d}_{ijk} N^-_k \\
\lbrack Q^A_i, N^-_j\rbrack &=& 0 \\ 
\{N^+_i, N^-_j\} &=& 2 \hat{d}_{ijk} Q^A_k \\
\{N^+_0, N^-_i\} &=& 2i Q^A_i
\end{array}
\eqno(5)
$$

\vfill\break
\noindent
{\bf 4. The Superconnection, Supercurvature and the Lagrangian} 

At this stage we set up the relevant superconnection \`{a} la Quillen,
as an ad hoc algorithm (we shall later discuss the possibility of
generating it from the matter fields' fiber bundle, by using a Connes-Lott
type of product base space). The superconnection will thus be given as
$$
\left(
\begin{array}{cc}
\Gamma^{\ \tilde b}_{\tilde a}\Sigma^{\tilde a}_{\ \tilde b} &
\Phi^{\ \ b_a\}}\Sigma^{\{a}_{\ b\}} \\
\Phi^{\ \ [b_{a]}}\Sigma^{[a}_{\ b]} & - \Gamma^{\ \tilde a}_{\tilde b}
\Sigma^{\tilde b}_{\ \tilde a} 
\end{array}
\right)
\eqno(6)
$$

The nonvanishing v.e.v. field $\phi(x)=\Phi^{+}_{0}$ will occupy the main
diagonal of quadrant II. This will also be the structure of the matrix
derivative $\delta$,
$$
\delta= \left(
\begin{array}{cc}
0 & i.1_{4\times 4} \\
0 & 0
\end{array}
\right)
\eqno(7)
$$

The resulting (generalized) curvature is then, 
$$
\hat{R}= R(G) +  + \{\Phi_{+},\Phi_{-}\} + D\Phi_{+} + D\Phi_{-} +
\delta\Phi^{-}
\eqno(8)
$$
where $\Phi_+, \Phi_-$ respectively denote the symmetric (in quadrant II)
and antisymmetric (in quadrant III) components of $\Phi_H$.
The first two terms arise for the 15 $\hat{R}^G$, the last three appear
for the 16 $\hat{R}^{H}$. In addition to its action on the Grassmann
algebra --replacing an n-form by a (4-n)-form --  the ${}^{*}$ duality
operator conjugates the supermatrix.  The $\hat{R}\wedge{}^{*}\hat{R}$
gauge Lagrangian will thus consist of the following terms  

(a) $R\wedge{}^{*}R$, the SKY Lagrangian $[21-23]$,

(b) $|\Phi^{-}|^{2}\phi^{2}$, the $\Phi^{-}$ mass term,
once $<0|\Phi^{+}_{0}|0>\neq 0$.

(c) $|\{\Phi^{-},\Phi^{+}\}|^{2}$, the quartic Higgs potential $V_{4}$.

(d) $(D\Phi_{+})^{2}$ the $\Phi_{+}$ kinetic energy and gauge interaction,

(e) $(D\Phi_{-})^{2}$, the $\Phi_{-}$ kinetic energy and gauge interaction,

(f)  $|\delta\Phi^{-}|^{2}$, the ``negative squared mass" term $V_{2}$,
triggering the spontaneous breakdown of local G symmetry, through
$\frac{\partial(V_2+V_4)}{\partial|\Phi^-|^2}=0$ 

(g) There is no $\{\Phi^{+},\Phi^{+}_{0}\}$ term, so that the 9
traceless components of $\Phi^{+}$ do not acquire mass.
Moreover, they become the longitudinal $G/F=SL(4,R)/SO(4)$-gauging
components of the connection, which acquire mass under the spontaneous
symmetry breakdown.
 
\vfill\break
\noindent
{\bf 5. The matter Lagrangian and Connes-Lott like Geometry}

We now discuss a Connes-Lott like derivation.
We stick to the {\it chiral} $Z_2$ grading, i.e. to the product
space $Z_2\otimes M^{3,1}=M_L\oplus M_R$ as base space. The matter fields
will consist of {\it world spinor manifields} [24-28],
the (spinorial) infinite-component representations of the double-coverings
$\overline{GL}(4,R)$ and $\overline{SL}(4,R)$, which, for several decades, were wrongly
assumed not to exist - in the General Relativity literature -- even though a
well-known algebraic theorem states that the topology of a Lie group is that
of its maximal compact subgroup,i.e.  $SO(4)\subset SL(4,R)$ and accordingly
$SU(2)\times SU(2)\subset \overline{SL}(4,R)$. We refer the reader to the relevant
literature, e.g. Chapter 4 and Appendix C of ref. [28]. 

The appropriate choice is the manifield based on
${\cal D}({1\over 2},0)\oplus{\cal D}(0,{1\over 2})$,
where $\cal D$ denotes the
$\overline{SL}(4,R)$ irreducible representation (applying the
{\it deunitarizing automorphism}${}^{[24]}$ $\cal A$) and
$({1\over 2},0), (0,{1\over 2})$
denote the lowest representations of the $SO(4)$ subgroup, here a nonunitary
representation of $SL(2,C)$, namely a Dirac spinor. We refer the reader to
the literature -- see figs. 3,4,5 of ref. [28] -- for a detailed discussion
of this field. Obviously, for a massless field,
$({1\over 2},0)$ and $(0,{1\over 2})$
respectively form the fibres over $M_{L}$ and $M_{R}$, with
$G=\overline{SL}(4,R)$ as a common structure group. The odd connection bridging
parallel-transport between points on the bundles constructed over
$M_L$ and $M_R$ has to contain (e.g. in a Yukawa-like term) a
$({1\over 2},{1\over 2})$ $\gamma^0$-supported scalar
$\Phi^{+}_{0}$ at least.
However, considering the structure of the manifield (see figs. 4,5
in ref. [28]), the $N^{+}$ generators with their $(1,1)$ action and the
$\Phi^{+}$ connections, are just what is needed to make a one-step bridging
along the fibre. The role of the $\Phi^{-}$, however, is rather unclear,
since the $N^{-}$ generators, with their $(1,0)\oplus (0,1)$ action,
act trivially. We have to be satisfied with the $N^{-}$ emerging from the
algebraic consistency of the supergroup structure, in going from the pure
Connes-Lott to its Marseilles-Mainz modification.

\end{document}